\documentclass[11pt]{article}
\usepackage[pdftex]{graphicx}
\usepackage[cmex10]{amsmath}
\usepackage{amsfonts}
\usepackage{stfloats}
\usepackage{fullpage}
\usepackage{color}
\usepackage{amsmath,amsfonts,amssymb}

\def\BM#1{\mbox{\boldmath$#1$}} 
\newcommand{\bfnu}{\BM{\nu}}
\newcommand{\bfx}{\BM{x}}
\newcommand{\bfy}{\BM{y}}
\newcommand{\bfd}{\BM{d}}

\newcommand{\real}{\mathbb{R}}
\newcommand{\hbfx}{\hat{\bfx}}
\newcommand{\E}{\mathbb{E}}
\newcommand{\R}{\mathbb{R}}

\newcommand{\calN}{{\mathcal N}}

\newcommand{\calP}{{\mathcal P}}

\newcommand{\var}{{\rm var\,}}

\def\iint{\int\kern-5pt\int\kern2pt}
\def\iiiint{\int\kern-5pt\int\kern-5pt\int\kern-5pt\int\kern2pt}

\def\tr{{}^{{\rm T}}}

\def\cond{\,\vert\,}
\def\calM{{\mathcal{M}}}

\def\mtrx#1#2{
  \left(
    \begin{array}{#1}
      #2
    \end{array}
  \right)}

\begin{document}

\title{A Bayesian approach to improving the Born approximation for
  inverse scattering with high contrast materials}

\author{J. P. Kaipio${}^{1}$, T. Huttunen${}^{2}$, T.
  Luostari${}^{2}$, T. L\"ahivaara${}^{2}$, and P. B. Monk${}^{3}$}

\date{
${}^{1}$University of Auckland, Department of Mathematics, Auckland, New Zealand\\
${}^{2}$University of Eastern Finland, Department of Applied Physics, Kuopio, Finland\\
${}^{3}$University of Delaware, Department of Mathematical Sciences, Newark DE 19716, USA}

\maketitle
\subsection*{{\bf Abstract}}

Time harmonic inverse scattering using accurate forward models is
often computationally expensive.  On the other hand, the use of
computationally efficient solvers, such as the Born approximation, may
fail if the targets do not satisfy the assumptions of the simplified
model.  In the Bayesian framework for inverse problems, one can
construct a statistical model for the errors that are induced when
approximate solvers are used, and hence increase the domain of
applicability of the approximate model.  In this paper, we investigate
the error structure that is induced by the Born approximation and show
that the Bayesian approximation error approach can be used to
partially recover from these errors.  In particular, we study the
model problem of reconstruction of the index of refraction of a
penetrable medium from measurements of the far field pattern of the
scattered wave.

\section{Introduction}

We investigate the classical model inverse problem of determining the
index of refraction of a bounded medium from measurements of the far
field pattern of the time harmonic scattered wave.  This is a
nonlinear ill-posed problem, for which several theoretical and
numerical methods have been proposed~\cite{colton+kress3}.  For
example, one can apply a least squares optimisation approach in which
the unknown refractive index is determined by fitting the measured
data using an optimisation algorithm applied to a suitable misfit
function (with appropriate regularisation terms).  This approach is
very flexible and often used, but has the disadvantage that it is
computationally intensive and may suffer from the presence of local
minima in the cost functional (see for example Hohage~\cite{hoh01} for
the use of fast integral solvers, and Bao et al. \cite{bao15} for
another approach to this problem).  In addition, regularisation
approaches do not provide statistically meaningful error estimates.

A popular alternative is to solve a linearised version of the problem.
In particular, in this paper we shall be concerned with the use of the
Born (or weak scattering) approximation.  For a review of
linearisation methods, see \cite[Chapter 8]{Devaney} and in the
context of radar scattering see \cite{Cheney+Borden}.  The Born
approximation underlies also some modern approaches to ultrasound
imaging (see \cite{Alberti17}). When applicable, these approaches are
very fast, but are potentially inaccurate if the true scatterer is not
weak.  Our numerical results in Section~\ref{sec:Results} will provide
examples of this inaccuracy.

Our aim is to increase the domain of applicability of inverse
scattering methods based on the Born approximation to high contrast
and multiple scattering cases.  We do this by using the Bayesian
Approximate Error (BAE) formulation of the inverse problem.  In
particular, errors that are related to using an approximate forward
solver (the Born approximation) are modelled by the BAE approach.
This is based on computing the approximate statistics of the modelling
related errors over a prior distribution for the index of refraction.
While the Bayesian inversion paradigm equipped with the approximation
error approach has proven to be a feasible combination for treating
different types of elliptic and parabolic problems with both simulated
and real data, this approach has not been applied to classical time
harmonic inverse scattering problems.

To describe the inverse problem in more detail, let us consider the
forward problem for time harmonic acoustic scattering from a bounded
penetrable obstacle.  Let $k$ denote the wave number of the acoustic
field.  In particular, if the temporal frequency of the sound is $f$
and if $\omega=2\pi f$ ($\omega$ is the angular frequency), then the
wave number $k$ is given by $k=\omega/c$ where $c$ is the speed of
sound in the background medium (i.e. outside the scattering object).

To simplify computations, we restrict our investigation to
$\mathbb{R}^2$.  Suppose we are given an incident plane wave
$u^i(\bfx,\bfd)=\exp(ik\bfx\cdot\bfd)$ with direction of propagation
$\bfd$, $|\bfd|=1$, and a possibly complex coefficient $n(\bfx)$ such
that $\Re(n(\bfx))>0$ and $\Im(n(\bfx))\geq 0$ for all $\bfx$ together
with the boundedness condition that the contrast
$m(\bfx):=1-n(\bfx)=0$ if $|\bfx|>R$ for some $R>0$.  In
electromagnetic applications $n$ represents the relative permittivity
of the medium, whereas for acoustic applications it denotes the square
of the refractive index.  Then the total field $u:=u(\bfx,\bfd)$ and
scattered field $u^s:=u^s(\bfx,\bfd)$ satisfy
\begin{eqnarray}
\Delta u+k^2n(\bfx)u&=&0\mbox{ in }\real^2,\label{helmholtz}\\
u&=&u^i+u^s\mbox{ in }\real^2,\label{uius}\\
r^{1/2}\left(\frac{\partial u^s}{\partial r}-iku^s\right)&\to&0 \mbox{ as } r:=|\bfx|\to\infty.\label{sommerfeld}
\end{eqnarray}
For rather general piecewise smooth $n\in L^\infty(\mathbb{R}^2)$,
well-posedness of this problem for $u\in H_{\rm loc}^1(\real^2)$ can
be proved (for details see \cite{wer62} and \cite{colton+kress3}).

It follows from the Sommerfeld radiation condition (\ref{sommerfeld})
that $u^s$ has an asymptotic expansion for large $|\bfx|$ as an
outgoing cylindrical wave: 
\[
u^s(\bfx,\bfd)=\frac{\exp(ik|\bfx|)}{\sqrt{|\bfx|}}\left(u_{\infty}(\hbfx,\bfd)+O\left(\frac{1}{|\bfx|}\right)\right),
\mbox{ as }|\bfx|\to\infty
\]
where $\hbfx=\bfx/|\bfx|$ is the observation direction (see equation
(3.86) of Colton and Kress \cite{colton+kress3}).  Given the incident
field and index of refraction, we can predict the far field pattern by
solving the well posed \emph{forward problem} for $u$ in some domain
including the scatterer based on (\ref{helmholtz})-(\ref{sommerfeld})
and then computing the far field pattern using the upcoming formula
(\ref{h-far}) or (\ref{h-farV2}).  In this paper, the far field
patterns will be used as data for the inverse problem of determining
$m$ (or equivalently $n$) which we describe next.

In particular, suppose now that $n$ is unknown (except that the
background value $n=1$ outside the bounded scatterer is given).  Let
$\Omega=\{\hbfx\;|\;|\hbfx|=1\}$ and suppose that we can measure an
approximation to $u_{\infty}(\hbfx_\ell,\bfd_j)$ for $N_w$ measurement
directions $\hbfx_\ell\in \Omega$, $1\leq \ell\leq N_w$, and incident
field directions $\bfd_j\in \Omega$, $1\leq j\leq N_w$.  In practice
we choose uniformly spaced directions on the unit circle and
$\bfd_j=\hat{\bfx}_j$.  Note that this is termed ``multistatic'' data
since we have assumed that the measurement and source directions can
be located independently of one-another.  Given this multistatic data
(at a single fixed frequency $k$), we seek to determine an
approximation to the unknown function $m$ (equivalently $n$).  This is
a non-linear ill-posed problem (see for example
\cite{hoh01,colton+kress3}).

In this paper, we shall use the Born approximation (described in
Section~\ref{born_intro}) as a fast approximate solver for the forward
problem and adopt a Bayesian framework for solving the inverse
problem. In particular, we use the so-called Bayesian approximation
error (BAE) approach to take into account the errors related to using
the Born approximation, as well as measurement errors.  The BAE will
be described in more detail in Section \ref{sec:BAE}, but for now we
note that the BAE is based on a normal (Gaussian) approximation for
the joint probability distribution of the primary unknowns and an
additive model for the approximation errors.  As a result, an affine
estimator is obtained, the computational complexity of which is
similar to a Tikhonov type regularised Born approximation.  The
Bayesian approach also allows us to compute (posterior) error
estimates.  Due to the structure of the BAE approximation, the error
estimates can be computed for a particular measurement setting before
any measurements are carried out. Furthermore, the error estimates
given by the approximation error approach are almost always feasible
and often only slightly larger than when accurate computational models
are employed \cite{jhuttunen2007b}.

The BAE approach has proven to be a computationally attractive
alternative to using computationally accurate forward solvers.  In
addition to coping with modelling errors such as finite element
discretisation, also errors due to using approximate physical models
have turned out to be feasible.  Model reduction and unknown
anisotropy structures in optical diffusion tomography were treated in
\cite{arridge2006,heino04,heino05a}.  The errors related to the
linearisation of the forward model of the optical tomography problem
were considered in \cite{tarvainen2010b}.  Missing boundary data in
the case of image processing and geophysical ERT/EIT were considered
in \cite{calvetti2005} and \cite{lehikoinen2007}, respectively.
Furthermore, overcoming errors in domain geometry was treated in
\cite{nissinen2008,nissinen2009b,nissinen2011}.  Also, in
\cite{nissinen2008,nissinen2009b,nissinen2011}, the problem of
recovery from simultaneous geometry errors {\em and} drastic model
reduction was found to be possible.  In \cite{tarvainen2010}, an
approximate physical model (diffusion model instead of the radiative
transfer model) was used for the forward problem while in
\cite{lahivaara14, lahivaara15} a poroelastic wave propagation forward
model is replaced with an elastic counterpart.  The discretisation
related errors in the context of ultrasound tomography were considered
in \cite{koponen2014}.  In \cite{kolehmainen2011}, an unknown
distributed parameter (scattering coefficient) was treated with the
approximation error approach.  In \cite{lahivaara14dg}, the unknown
subsurface material properties are marginalised using the BAE in the
context of electromagnetic wave propagation (ground-penetrating
radar).

The closest papers to ours are \cite{lahivaara14dg} and
\cite{koponen2014}.  In \cite{lahivaara14dg}, Maxwell's equations were
used in the time domain, whereas we are working in the frequency
domain.  Also the concern in that paper was to handle random
background physical parameters and seek the position of objects, while
we wish to obtain quantitative estimates of the refractive index. The
work of \cite{koponen2014} is also in the time domain, and both papers
use a forward model that is non-linear with respect to the physical
parameters. Our study is the first in the frequency domain, and we use
an approximate fast linear solver.

Our BAE approach uses the solution of many accurate forward solves of
the problem (with randomly chosen data) to predict the error
statistics for the Born approximation (for example 3,000 forward
problems were used when generating our numerical results in
Section~\ref{sec:Results}). This is called the \emph{training} phase
of the algorithm. It is separate from solving the inverse problem, and
can be done in parallel. The data from these solutions is then used to
correct the Born approximation.  The \emph{online} phase when the BAE
corrected Born approximation is used is very rapid (as is the usual
Born approximation). There is no need for further solution of the
accurate model in the online phase. Thus the method is most suitable
for situations (for example process monitoring) in which it is needed
to solve the inverse problem rapidly many times.  If just a single
inverse problem needs to be solved, then a more standard non-linear
optimisation approach might be more appropriate, although the
programming of such a method, solving appropriate forward and adjoint
problems at each optimisation step, is more complex than for the BAE
scheme.

The rest of the paper is organised as follows: In
Section~\ref{born_intro} we give some details and comments on the Born
approximation.  In Section~\ref{sec:Bayes}, we give a brief overview
of the Bayesian framework for inverse problems.  In
Section~\ref{sec:BAE}, we outline the Bayesian approximation error
approach and derive the general form for the estimator.  In
Sections~\ref{sec:ForwardModels} and \ref{sec:Priors}, we define the
related computational forward models and the prior models used in this
paper, respectively.  In Section~\ref{sec:Results}, we consider
several computational examples.

\section{Born approximation}
\label{born_intro}

The BAE approach to inverse problems can be used to allow for fast
approximate forward solvers.  In this paper, such a solver is provided
by the Born approximation which we now describe.

One way to prove the existence of a solution to
(\ref{helmholtz})-(\ref{sommerfeld}) is by showing that this
problem is equivalent to solving the Lippmann-Schwinger equation
satisfied by $u\in H^1_{\rm{}loc}(\mathbb{R}^2)$:
\begin{equation}
  u(\bfx)=u^i(\bfx)-k^2\int_D \frac{i}{4}H^{(1)}_0(k|\bfx-\bfy|)m(\bfy)u(\bfy)\,dA(\bfy),
  \qquad\forall \bfx\in\real^2\label{LS}
\end{equation}
where $H_0^{(1)}$ is the Hankel function of first kind and order zero,
and  and $D$ contains the support of
$m$ in its interior (see \cite{hoh01}).

The far field pattern can then be computed by applying an asymptotic
analysis to the Lippmann-Schwinger equation (\ref{LS}) for large
$|\bfx|$ to obtain the identity
\begin{equation}
u_{\infty}(\hbfx,\bfd)=-\frac{k^2\exp(i\pi/4)}{\sqrt{8\pi k}}\int_{D}\exp(-ik\hbfx\cdot\bfy)m(\bfy)
u(\bfy,\bfd)\,dA(\bfy)
\label{h-far}
\end{equation}
where $|\hbfx|=1$, $m(\bfy)=1-n(\bfy)$. An equivalent way to obtain
the far field pattern is by enclosing $D$ by a simple closed curve $C$
and evaluating the following integral (equation (3.87) in
\cite{colton+kress3})
\begin{eqnarray}
\lefteqn{u_{\infty}(\hbfx,\bfd) = \frac{\exp(i\pi/4)}{\sqrt{8\pi k}}\times} \nonumber\\
\displaystyle&&\int_{C}\left(
	u(\bfy,\bfd)\frac{\partial\exp(-ik\hbfx\cdot \bfy)}{\partial \bfnu} 
		- \frac{\partial u}{\partial \bfnu}(\bfy,\bfd) \exp(-ik\hbfx\cdot \bfy)
\right)\,dA(\bfy)
\label{h-farV2}
\end{eqnarray}
where $\bfnu$ denotes the outward normal on $C$. We use this
representation in the training phase of BAE to compute the far field
pattern of a known scatterer using the solution obtained from an
accurate but computationally more demanding finite element forward
solver in a neighbourhood of $D$.  This solver is described in
Section~\ref{sec:ForwardModels}.
 
The first term of the Neumann series for solving (\ref{LS}) provides
 the Born approximation of the field $u$:
\begin{equation}
  u(\bfx)\approx u^i(\bfx)-k^2\int_D \frac{i}{4}H^{(1)}_0(k|\bfx-\bfy|)m(\bfy)u^i(\bfy)\,dA(\bfy),
  \qquad\forall \bfx\in\real^2\label{BS}
\end{equation}
In the same way, the Born approximation for the far field pattern
(\ref{h-far}) is obtained by replacing $u(\bfy,\bfd)$ by
$u^i(\bfy,\bfd)$ in (\ref{h-far}) which we consider as an operator
from the contrast $m$ to an approximate far field pattern $F(m)$.
This yields
\begin{equation}
u_{\infty}(\hbfx,\bfd)\approx F(m):=-\frac{k^2\exp(i\pi/4)}{\sqrt{8\pi k}}\int_{D}
	\exp(ik(\bfd-\hbfx)\cdot\bfy)m(\bfy)\,dA(\bfy).\label{born}
\end{equation}
The Born approximation is justified for low wave number or small
contrast. Suppose $D$ is contained in a ball of radius $a$, then the
(\ref{BS}) will be accurate if $(ka)^2\Vert m\Vert_{L^{\infty}(D)}$ is
sufficiently small. If the Neumann series for (\ref{BS}) does not
converge (e.g at high frequency or for large contrast) the Born
approximation may still be useful to obtain qualitative information
about the scatterer but can no longer be expected to give quantitative
estimates of $m$.  In particular, for problems in which the
inhomogeneities in the index of refraction are on the same scale as
the wavelength, or the contrast in the index of refraction are high,
accurate forward solvers usually have to be employed.  In such cases,
the computational complexity of optimisation approaches to the inverse
problem can often turn out to be formidable.

As mentioned above, the Born approximation underlies many fast
inversion schemes but neglects multiple reflections.  It also becomes
less accurate as $m$ increases, or for large $k$ \cite{colton+kress3}
as noted above.  These deficiencies can introduce phantoms in the
reconstructions \cite{sim07,Cheney+Borden} or completely destroy the
reconstruction.

Our goal is to use the Born approximation as a fast forward solver
even when it provides a very poor approximation of the far field
pattern.  In this paper, the Bayesian approximation error approach
will be used to correct for the solver related errors.


\section{Approximate marginalisation: Bayesian Approximation Error approach}

While the Born approximation allows for linear computational
reconstruction algorithms based on deterministic regularisation, such
methods are not able to provide statistically meaningful error
estimates.  On the other hand, while the Bayesian framework allows for
such error estimates, the overall model needs to accommodate all types
of errors, not merely the additive measurement errors.  When employing
approximate forward models as part of the inversion procedure, we also
need to include the related approximation/modelling errors.  The
structure of such errors is usually analytically intractable.

The Bayesian approximation error approach carries out a normal
approximation for such errors which allows for an efficient way to
embed these errors in the posterior model which expresses the
uncertainty of the refractive index given the measurements.  In
statistical terminology, this embedding is referred to as
marginalisation.

\subsection{Bayesian framework for inverse problems}
\label{sec:Bayes}
In the Bayesian framework for inverse problems, the measurements and
{\em all} unknowns, that is, both the interesting and uninteresting
unknowns, are modelled and treated as random variables, see
\cite{kaipio2005, tarantola2004,calvetti2007}.  In the context of this
paper, let $Y\in\R^M$ be a vector of measurements (real and imaginary
parts of the far field patterns for all combinations of measurement
and incident directions) and $m\in\R^N$ a vector of the interesting
unknowns (average values of the contrast pixel by pixel in the image)
and $\xi$ be uninteresting unknowns (for example measurement error and
modelling error).  In this paper, we take the contrast $m$ defining
the scatterer to be real-valued and, furthermore, the $M=2N_w^2$
real-valued measurements in $Y$ are the real and imaginary parts
corresponding to the $N_w^2$ far field measurements ordered as a
vector.  We construct a statistical model for the joint distribution
$\pi(Y,m,\xi)$ and derive the respective (conditional) {\em posterior
  distribution} $\pi(m\cond Y)$ which is a model for the uncertainty
in the random variables $m$ given the measurements $Y$, and in which
the uninteresting unknowns $\xi$ have been marginalised.  The models
$\pi(Y,m,\xi)$ and $\pi(m\cond Y)$ are usually constructed via Bayes'
formula
\[
\pi(Y,m,\xi) = \pi(Y\cond m,\xi)\pi(m,\xi) = \pi(m,\xi\cond Y)\pi(Y)
\]
whence (since once the measurements are carried out, $\pi(Y)$ is a
fixed although unknown real number) we formally obtain the posterior
distribution (density)
\[
\pi(m\cond Y) = \int \pi(m,\xi\cond Y)\, d\xi \propto \pi(Y\cond m)\pi(m)
\]
where the marginalisation takes the form of integration, and the two
densities on the right hand side are called the {\em likelihood
  density} and the {\em prior density}, and both these densities have
always to be considered as models only.

In the context of inverse problems, the construction of the likelihood
density involves the modelling of the forward problem (noiseless
measurements) when all related variables and parameters are known, and
a model for the observation errors and other uninteresting unknowns,
that is $\xi$.  The prior model should have the property that
$\pi(m')/\pi(m'')\gg1$ for all expected solutions $m'$ and unexpected
solutions $m''$.  For more comprehensive and systematic discussion,
see any of \cite{kaipio2005, tarantola2004,calvetti2007} or general
references in Bayesian statistics, such as
\cite{berger1980,robert2004,box1992}.

Once the posterior model has been derived, all questions that have
been posed in terms of probabilities, can be answered.  Typically,
when the number of interesting unknowns is $N>2$, point estimates are
computed, such as the maximum a posteriori and conditional mean
estimates
\begin{eqnarray*}
m_{\rm MAP} &=& \arg\max_{m} \,\pi(m\cond Y) \\
m_{\rm CM} &=& \E(m\cond Y) 
\end{eqnarray*}
as well as spread estimates such as the posterior covariance
$\Gamma_{m\vert Y}$, marginal posteriors $\var(m_\ell \cond Y)$, as
well as probabilities of events, such as $\calP(m_\ell\in(m_{-},m^{-})
\cond Y)$, that is, the probability that $m_\ell$ is between the two
arbitrary numbers $m_{-}$ and $m^{-}$, given $Y$.

Computing these answers may, however, be a complex task in the case of
inverse problems involving usually excessively time consuming Markov
chain Monte Carlo stochastic simulation.  When solving inverse
problems with limited computational resources, more or less severe
approximations have to be carried out which may lead to inaccurate or
even infeasible estimates unless the approximations are themselves
modelled \cite{kaipio2005,kaipio2007}.

In large dimensional limited resource problems, (possibly highly
approximate) linear forward models and a normal (Gaussian) model for
all related random variables may be the only computationally feasible
overall model since most of the computations can be carried out before
the measurement process.  Furthermore, the MAP and CM estimates would
then be (linear) affine functions of the measurements and the most
important spread estimates would not depend on the data and
computation involves linear algebra only.  This applies to the problem
in this paper too.  In the following, we consider such an
approximation and show how the related approximation/modelling errors
can be taken into account and marginalised.

\subsection{Bayesian approximation error approach}
\label{sec:BAE}

We now give a brief overview of the BAE approach to inverse problems.
We assume that the forward model is the only source for the
approximation and model errors.  For the general formulation of more
complex cases including auxiliary coefficients, unknown boundary data
etc, see \cite{kolehmainen2011, kaipio2013}.
 
Let $\bar F$ denote a computationally accurate forward model (for
example, given by a full, well resolved, finite element method) such
that
$$ Y =  \bar F(\bar m) + e \in\R^N$$ 
where $\bar m$ is an accurate approximation of $m$.  Here the the
associated modelling and measurement error $e$ is unknown but is
assumed to be mutually independent with $m$. Let $\pi(m,e)$ be a model
for the joint distribution of the unknowns.

We then approximate the accurate representation of the primary unknown
$\bar m$ by $m = P\bar m$ where $P$ is typically a projection
operator.  We identify $m = P\bar m$ with its coordinates in the
associated basis when applicable.

In the approximation error approach, we proceed as follows.  Instead
of using the computationally accurate nonlinear forward model $\bar m
\mapsto \bar F(\bar m)$, we use the Born approximation as a
computationally rapid but approximate forward model denoted as before
by $ m \mapsto F(m).  $ Thus, we write the measurement model in the
form
\begin{equation}
Y = \bar F(\bar m) + e 
  = F(m) + \varepsilon + e
   \label{eq:AEMod}
\end{equation}
where we define the {\em approximation error} $\varepsilon = \bar
F(\bar m) - F(P\bar m)$.  Thus, the approximation error is the
discrepancy of predictions of the measurements (given the best choice
of unknowns) when using the accurate model $\bar F(\bar m)$ and the
approximate model $F(P\bar m)$.

We employ approximate joint distributions and therefore consider
$\pi(\varepsilon\cond \bar m)$ without any special structure.  As a
first step, we approximate $\bar m\approx P\bar m$ and thus
$\pi(\varepsilon \cond \bar m) \approx \pi(\varepsilon \cond m)$.
This means that we assume that the model predictions and thus the
approximation error is essentially the same for $\bar m$ as $m = P\bar
m$.  This approximation is valid unless a very crude low dimensional
representation is employed for $m$.

The Bayesian approximation error approach deals with the approximate
marginalisation in the computation of the (approximate) likelihood
model $\pi(Y\cond m)$.  It was shown in \cite{kolehmainen2011,
  kaipio2013} that the exact marginalisation over the measurement and
additive errors leads to
\[
\pi(Y \cond m) = \int \pi_e(Y - F(m) - \varepsilon) 
      \pi_{\varepsilon\vert m}(\varepsilon\cond m)\,d\varepsilon
\]
At this stage, in the BAE, both $\pi_e$ and $\pi_{\varepsilon\vert m}$
are approximated with normal distributions. 

Let the normal approximation for the joint density $\pi(\varepsilon,m)$ be
\begin{equation*}
  \pi(\varepsilon,m) \propto \exp\left\{-\frac12 
    \mtrx{c}{\varepsilon - \varepsilon_\ast \\m - m_\ast}^{\rm T}
    \mtrx{cc}{\Gamma_{\varepsilon\varepsilon} & \Gamma_{\varepsilon m}\\
      \Gamma_{m\varepsilon} & \Gamma_{mm}}^{-1}
    \mtrx{c}{\varepsilon - \varepsilon_\ast \\m - m_\ast}
  \right\}
\end{equation*}
Thus, denoting the multivariate normal distribution
with mean $\mu$ and covariance $\Gamma$ by $\calN(\mu,\Gamma)$, we write
\[
e \sim \calN(e_\ast,\Gamma_e),\qquad
\varepsilon\cond m \sim \calN(\varepsilon_{\ast\vert m},\Gamma_{\varepsilon\vert m})
\]
where
\begin{eqnarray*}
  \varepsilon_{\ast\vert m} &=& \varepsilon_\ast + \Gamma_{\varepsilon m}\Gamma_{mm}^{-1} 
  (m - m_\ast),\\
  \Gamma_{\varepsilon\vert m} &=& \Gamma_{\varepsilon\varepsilon}
  - \Gamma_{\varepsilon  m}\Gamma_{mm}^{-1}\Gamma_{m\varepsilon}.
\end{eqnarray*}
Here the subscript $*$ denotes the mean of the relevant quantity.
Define the normal random variable $\nu$ so that
$\nu\cond m = e + \varepsilon\cond m$ then
\[
\nu\cond m \sim \calN(\nu_{\ast\vert m},\Gamma_{\nu\vert m}) 
\]
where
\begin{eqnarray}
  \label{eq:nutilde}
  \nu_{\ast\vert m} &=& e_\ast + \varepsilon_\ast 
  + \Gamma_{\varepsilon m}\Gamma_{mm}^{-1} (m - m_\ast), \\
  \Gamma_{\nu\vert m} &=& \Gamma_e + \Gamma_{\varepsilon\varepsilon} - 
  \Gamma_{\varepsilon m} \Gamma_{mm}^{-1}  \Gamma_{m\varepsilon}
\end{eqnarray}
Following \cite{kolehmainen2011, kaipio2013}, we obtain for the approximate likelihood distribution 
\[
\pi(Y\cond m \sim)=\calN(Y - F(m) - \nu_{\ast\vert
  m},\Gamma_{\nu\vert m})
\]

Since we are after computational efficiency, a normal approximation
for the prior model is also conventionally used:
\begin{equation}
  \label{eq:approxprior}
  m \sim \calN(m_\ast,\Gamma_{mm})
\end{equation}
The approximation for the posterior distribution can thus be written as
\[
\pi(m\cond Y) \propto \pi(Y\cond m)\pi(m)
\propto \exp\left( -\frac12 V(m\cond Y) \right)
\]
where $V(m\cond Y)$ is the posterior potential that can be written in
the form 
\begin{equation}
  \label{eq:PostPotential}
  V(m\cond Y) = \Vert L_{\nu\vert m} (Y - F(m) - \nu_{\ast\vert m})\Vert^2 + 
  \Vert L_m (m - m_\ast) \Vert^2 
\end{equation}
where $\Gamma_{\nu\cond m}^{-1} = L_{\nu\vert m}\tr L_{\nu\vert m}$
and $\Gamma_{mm}^{-1} = L_m\tr L_m$.  For example, the approximate
maximum a posteriori estimate (which in this case coincides with the
conditional mean estimate) is then obtained by computing the minimizer
of (\ref{eq:PostPotential}) and the approximate posterior covariance
takes the form
\begin{equation}
\label{eq:postcov}
\Gamma_{m\vert Y} = \left( \tilde{F}\tr \Gamma_{\nu\vert m}^{-1} \tilde{F} + \Gamma_{mm}^{-1} \right)^{-1}
\end{equation}
where $\tilde{F}=F+\Gamma_{\varepsilon m} \Gamma_{mm}^{-1}$.  A
further approximation, that is referred to as the enhanced error
model, is obtained by setting $\Gamma_{\varepsilon m} = 0$.  This
further approximation is more stable with respect to the numerical
approximation of $ \Gamma_{\nu\vert m}$.

A few points are to be noted.  First, if the prior model is assumed to
be a feasible model for a class of unknowns $m\in\calM$, the matrices
and vectors $(L_m,L_{\nu\vert m},m_\ast,\nu_{\ast\vert m})$ are the
same for $\calM$.  The computation of these matrices and vectors can
be a significant undertaking, but needs to be carried out only once,
and prior to carrying out the actual measurements and the inversion
process (an offline training phase).  Furthermore, since the Born
approximation is a linear map, the posterior mean estimate is an
affine map of the measurements and we can write
\[
\E(m\cond Y) = BY + c
\]
where $B$ and $c$ can be precomputed.  The same applies for the
posterior covariance.




\subsection{The forward models in this paper}
\label{sec:ForwardModels}

The forward scattering problem (\ref{helmholtz}-\ref{sommerfeld}) can
be written in the terms of the scattered field as the problem of
determining $u^s\in H^1_{\rm{}loc}(\mathbb{R}^2)$ such that
\[
\Delta u^s + k^2n(\hbfx) u^s = k^2 (1-n(\hbfx)) u^i
\]
together with the (asymptotic) Sommerfeld radiation condition.  This
can be numerically approximated, for example, using the finite element
method (FEM) in a computational domain $\bar\Omega$ which is somewhat
larger than the support of the scatterer by employing the perfectly
matched layer (PML) outside $\bar\Omega$ to truncate the computational
domain.  The integral (\ref{h-farV2}) can then be used to compute the
far field pattern based on the FEM solution on a simple closed curve
$C$ containing $D$ ($C=\partial D$ is allowed for simple scatterers)
as described in \cite{sul98}.  We use this approach to provide the
computationally accurate forward model $\bar F(\bar m)$.  The
representation for $\bar m$ is in terms of the FE mesh: in particular,
$\bar m$ is approximated using a piecewise constant basis on the same
mesh as is used by the FEM.  Further details of the FEM implementation
will be given in Section~\ref{sec:Results}.

The Born approximation $F(m)$ is computed by approximating $m$ using a
piecewise constant basis on a uniform rectangular grid that covers
$D$.  The integral (\ref{born}) can then be computed analytically.

\subsection{Prior models and approximation error statistics}
\label{sec:Priors}

We employ two different prior models: one for the construction of the
model for the joint statistics of the approximation errors
$\varepsilon$ and the other for the primary unknown $m$.  In this
paper, we use a model $\bar\pi(\bar m)$ in which the domain $D$
contains $1-3$ ellipses whose number, locations, principal axes and
refractive index values are random, with details given in
Section~\ref{sec:Results}.  Four draws from $\bar\pi(\bar m)$ are
shown in Section~\ref{sec:Results} in Fig.~\ref{fig:samples}.

Once $p$ draws $\bar m^{(\ell)},\ \ell=1,\dots,p$ from this prior have
been drawn, the respective approximation errors $\varepsilon^{(\ell)}
= \bar F(\bar m^{(\ell)}) - F(P\bar m^{(\ell)})$ are then computed.
Then, the (joint) sample means and sample covariances are computed and
employed in (\ref{eq:nutilde}-\ref{eq:postcov}). This training is done
once before solving any inverse problems, and the results are then
used without modification for all the reconstructions shown in the
next section.

The other prior model $\pi(m)$ is the one used in the inversion. We
take this model to be a normal approximation for the unknown, since we
aim for an efficient inverse solver.  For this paper, we use a normal
prior $\pi(m) = \calN(m_\ast,\Gamma_{mm})$ where $m_\ast = 0$ and
$\Gamma_{mm}$ is a homogeneous isotropic Ornstein-Uhlenbeck covariance
with characteristic length $\lambda$. The prior standard deviations
are set to $\sigma_{m,k} = 0.4$ for all $k=1,\dots,N$.

The main reason to test any Bayesian approach with two different prior
models is to assess the robustness of the overall approach, and help
avoid ``inverse crimes''.  In the Bayesian framework, an inverse crime
might arise when using a prior model and considering true targets that
are draws from this (prior) model.  This could be interpreted as
having an exact information on the actual distribution of the unknown.
For this reason, the test targets are often chosen not to be draws
from the prior model.  To get around this concern in this manuscript,
we proceeded as follows.  We use:
\begin{enumerate}
\item A prior model consisting of a random number of randomized
  ellipsoids.  This model is not a random field and definitely not a
  normal field.  This can be taken as ``our best guess what the
  scatterers could look like''.  This model, however, is not a
  feasible one for the inversion in this case: the (random) ellipsoid
  model could easily be implemented but the resulting computations
  would involve Markov chain Monte Carlo algorithm which are
  notoriously slow.  The motivation of the present manuscript is to
  use approximate models to allow for fast computations in the first
  place.
\item The prior model for the unknowns that is used in the inversion
  is taken to be a normal random field.  Draws from this model do not
  generally resemble ellipsoids and definitely not a resonator.  In
  the construction of the normal prior model (the mean and the
  covariance matrix), the related standard deviations are made to
  correspond to the extrema of the index of refraction over the
  scatterer.  Such approximate information is readily available in any
  practical application.
\item In the first two numerical tests in Section~\ref{sec:Results},
  the actual targets conformed to the above prior (one or several
  ellipsoids).  In the last test in Section~\ref{case3}, the resonator
  does not conform to the above model for scatterers.  This is the
  most difficult case for the BAE corrected Born approximation.
\end{enumerate}

\section{Numerical experiments}
\label{sec:Results}

\subsection{Computational models}

In this section, we show the results in terms of the refractive index
(squared) $n$ instead of the contrast $m$.  For the background medium
we set $\bar n=1$.  The background wave number is $k = 15$.  We
consider multistatic far field data with $N_w=61$ angularly
equi-distributed plane waves and with $N_w=61$ far field pattern
measurement directions giving us $N = 2 N_w^2 = 7442$ measurements
(real and imaginary parts) of the far field pattern.

For both the accurate forward model $\bar F$ and the Born
approximation $F$, we set the computational domain to be $\Omega =
[-1.0,\ 1.0]\times[-1.0,\ 1.0]\in\R^2$.  This choice of $\Omega$
represents a priori known data on the location of the support of the
scatterers.  For $\bar F$, we employ the FEM with meshing that is
adapted to the geometry of the respective draws $\bar m^{(\ell)}$ from
$\bar \pi(\bar m)$.  Third order basis functions are used for the FEM
approximation.  For the computation of the actual measurements, the
maximum elements size $h_{\max}=0.03$ (elements per wavelength
$\lambda/h_{\max}\approx 14$).  For the computation of the draws for
BAE, we set $h_{\max}=0.05$ corresponding to $\lambda/h_{\max}\approx
8.4$.  The functions $\bar n(\bfx)$ is taken to be piecewise constant.
The Cartesian PML layer has depth 0.5, that is, the entire
computational domain for the FEM solver is $[-1.5,\ 1.5]\times[-1.5,\
1.5]$.  Once the accurate solution for the field $u$ has been
computed, Eqn.~(\ref{h-farV2}) is used to compute the far field
pattern for all incident waves using the approximation of the normal
derivative term as in as in \cite{sul98}.  Mutually independent zero
mean Gaussian noise (STD 3\% of the peak value of the measurements) is
then added to both real and imaginary parts of the far field data.

For the Born approximation forward model, the function $n(\bfx)$ is
again taken to be piecewise constant but now on a uniform square grid
of 2500 pixels over the computational domain $\Omega = [-1.0,\
1.0]\times[-1.0,\ 1.0]$.  The Born forward map (\ref{born}) is then
computed analytically.

\subsection{BAE samples}

\begin{figure}[ht]
 \begin{center}
  \includegraphics[width=0.8\textwidth]{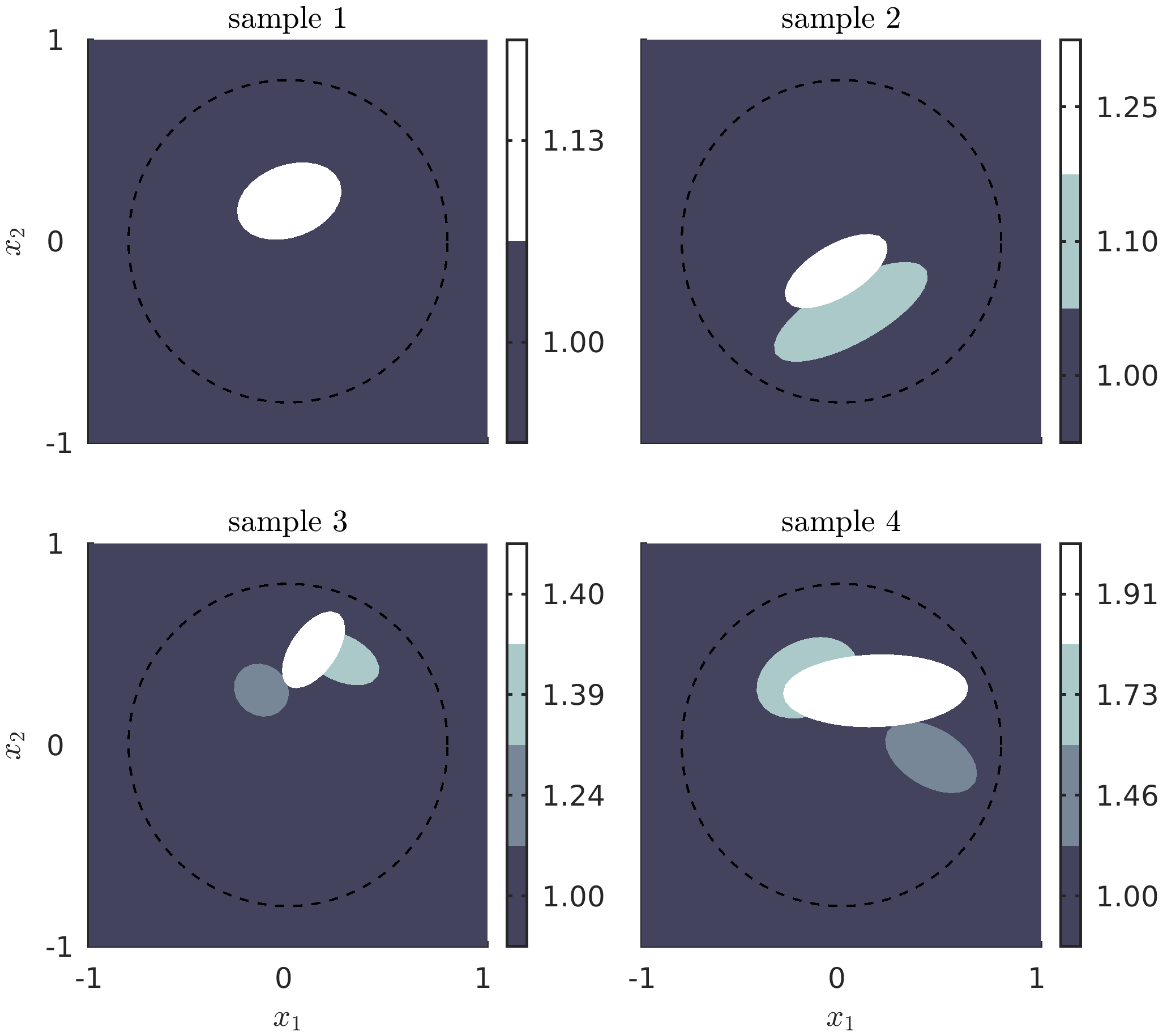}
 \end{center}
 \caption{Four samples from $\bar\pi(\bar n)$. The dashed line denotes
   the boundary $\partial D$.} \label{fig:samples}
\end{figure}

A total of 3000 samples were used to generate error statistics in BAE.
The choice of training data in practice would be made on the basis a
priori information about the general extent and composition of the
scatterer(s).  For the numerical experiments here, we choose the
training data to consist of a scatterer composed of from one to three
ellipses with randomised shape, location, orientation and material
properties in $D$.  The domain $D$ is chosen to be a disc of radius
0.8.  Knowledge of the approximate location and extent of the
scatterer is essential a priori information for the method.  Figure
\ref{fig:samples} shows four draws used for the computation of the BAE
statistics.  For all draws $\bar n(\bfx)^{(\ell)}$, the approximation
errors $\varepsilon^{(\ell)} = \bar F(1-\bar n(\bfx)^{(\ell)}) -
F(P(1-\bar n(\bfx)^{(\ell)}) )$ are computed.  Subsequently, the joint
mean and covariance of $(n,\varepsilon)$ are computed.

\begin{figure}[ht]
 \begin{center}
  \includegraphics[width=0.8\textwidth]{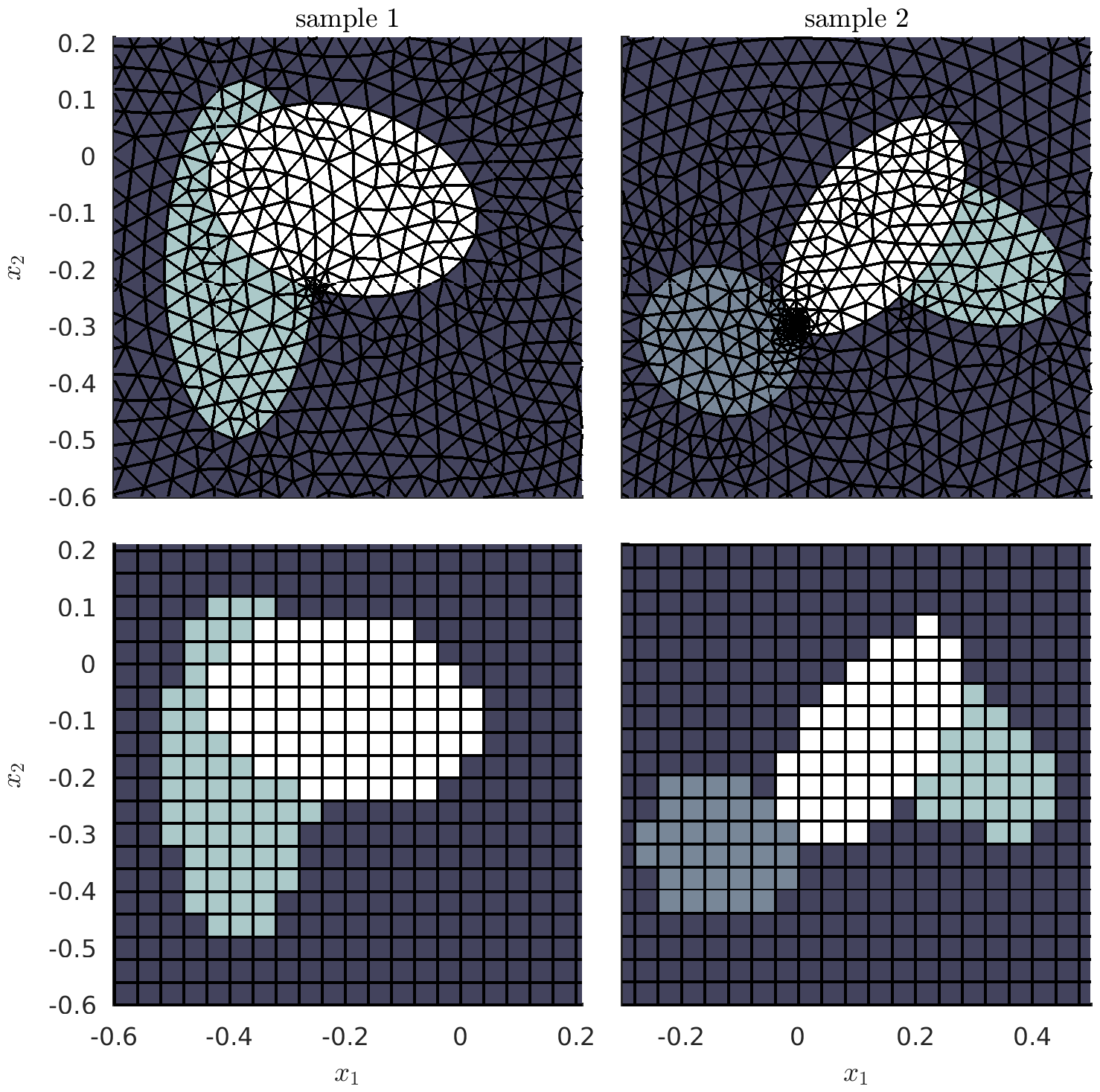}
 \end{center}
 \caption{Two examples of draws from $\bar\pi(\bar n(\bfx))$ and the
   respective FEM meshes (top) and the corresponding Born grid and the
   projections $n = P\bar n$ (bottom).  Left hand example: 10110
   finite elements, 5176 vertices, $h_{\max}=0.06377$,
   $h_{\min}=0.00445$.  Right hand example: 10362 finite elements,
   5302 vertices, $h_{\max}=0.06377$, $h_{\min}=0.00271$. The Born
   grid is always a uniform $50\times 50$ grid.  }\label{fig:grids}
\end{figure}

Figure \ref{fig:grids} shows two examples of the computational grids.
The FEM model generates a unique mesh for each draw from $\bar\pi(\bar
n(\bfx))$.  The Born approximation uses a fixed number of pixels to
represent $n(\bfx)$.

\subsection{Case 1: A  single ellipsoidal scatterer}
 
For each case, we compute the estimates of $n$ using the Born
approximation with and without the Bayesian approximation error
approach. In addition, with the BAE, we compute the estimates both
with the full and the enhanced error models.

\begin{figure}[ht]
 \begin{center}
  \includegraphics[width=0.8\textwidth]{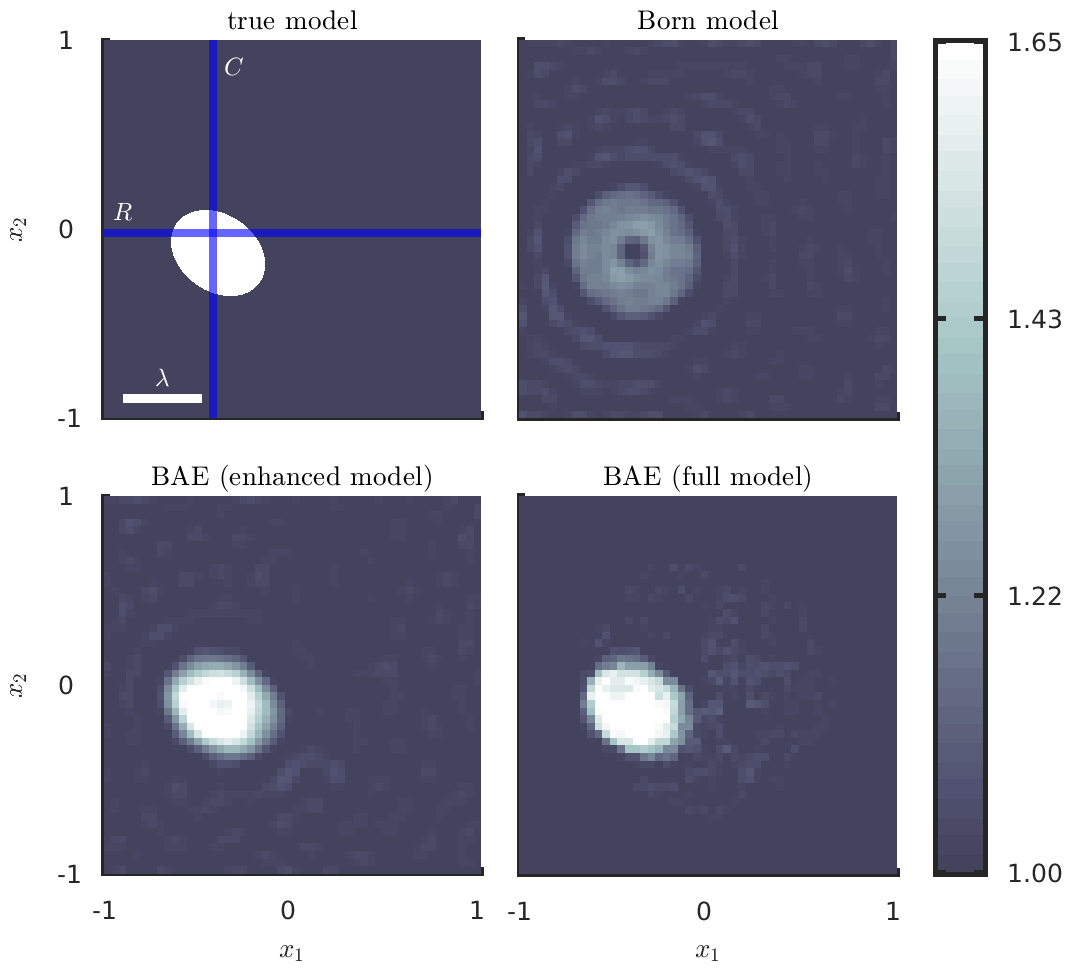}
 
  \caption{The reconstruction (estimates) of a single ellipsoidal
    scatterer with $\bar n = 1.65$.  Top left: the true $\bar n$. Top
    right: the standard Born approximation. Bottom left: the enhanced
    error model.  Bottom right: the full BAE model.  The wavelength
    $\lambda$ is indicated at the bottom left corner of the true
    model.  The actual $\bar n$ and the reconstructions along the
    cross sections $R$ and $C$ are shown in
    Fig.~\ref{fig:onescatterer_estimate_line}. }
     \label{fig:onescatterer_estimate}
\end{center}\end{figure}

In this example, we have a single elliptic target with $\bar n = 1.65$
inside the ellipse.  The actual $\bar n$, and results for the Born,
the full BAE and the enhanced error estimates are shown in
Fig.~\ref{fig:onescatterer_estimate}. The contrast is relatively high
and the Born approximation is inaccurate.  This is seen in the
estimate with an exaggerated size, with ringing and underestimated
values of $\bar n$ inside the target (see top right panel of
Fig.~\ref{fig:onescatterer_estimate}).  The location of the target is
approximately correct.

Both BAE estimates, on the other hand, estimate the size, the location
and the values well.  There is no significant ringing effect which
would be due to the underestimation of the values within the target.
Also the slightly elliptic shape of the target is roughly
recognisable.

\begin{figure}[ht]
 \begin{center}
  \includegraphics[width=0.8\textwidth]{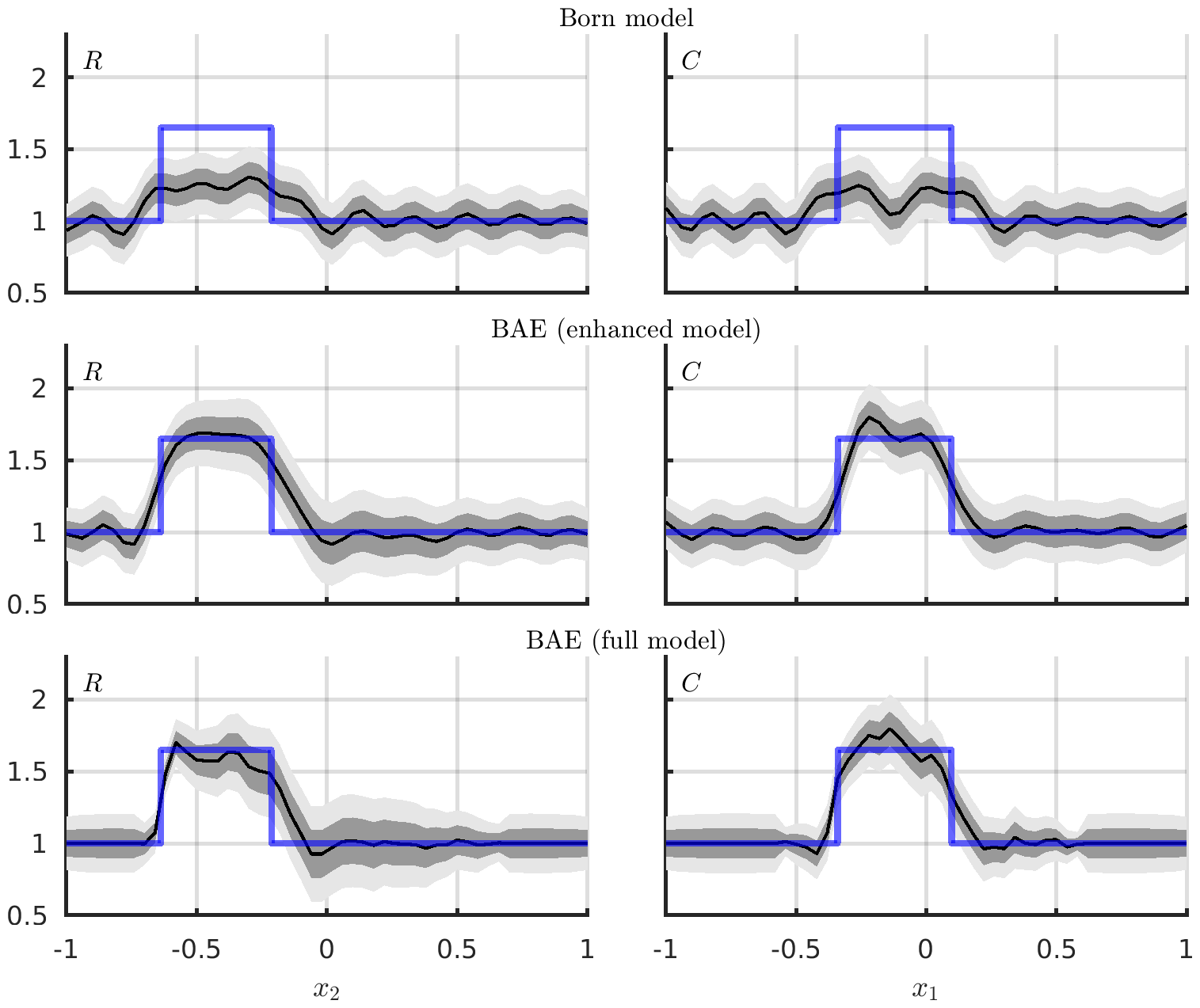}
 \end{center}
 \caption{Estimates along the lines $R$ (left column) and $C$ (right
   column) for the single scatterer of Fig.
   \ref{fig:onescatterer_estimate}. Transparent blue lines show the
   true model while black line shows the MAP-estimate.  Dark grey
   denotes $\pm$1 standard deviation while light grey denotes $\pm$2
   standard deviations.}
 \label{fig:onescatterer_estimate_line}
\end{figure}

Two cross sections of the actual $\bar n$ and the three estimates are
shown in Fig.~\ref{fig:onescatterer_estimate_line}. We also show the
$\pm1$ and $\pm2$ posterior standard deviations along these cross
sections.  Qualitatively, the estimates exhibit the same
characteristics as Fig.~\ref{fig:onescatterer_estimate}.  In addition,
most importantly, the BAE error estimates are useful in that the
actual cross sections are within two posterior standard deviations of
the reconstructions.  This does not apply to the Born estimate.  While
one of the most significant assets of the Bayesian framework is that
one can obtain posterior error estimates, this example shows that if
any errors or uncertainties are not modelled (as is the case for the
basic Born approximation shown in the top row of
Fig.~\ref{fig:onescatterer_estimate_line}), the posterior error is
typically underestimated.

\clearpage

%
\subsection{Case 2: Three ellipsoidal scatterers}
\begin{figure}[ht]
 \begin{center}
  \includegraphics[width=0.8\textwidth]{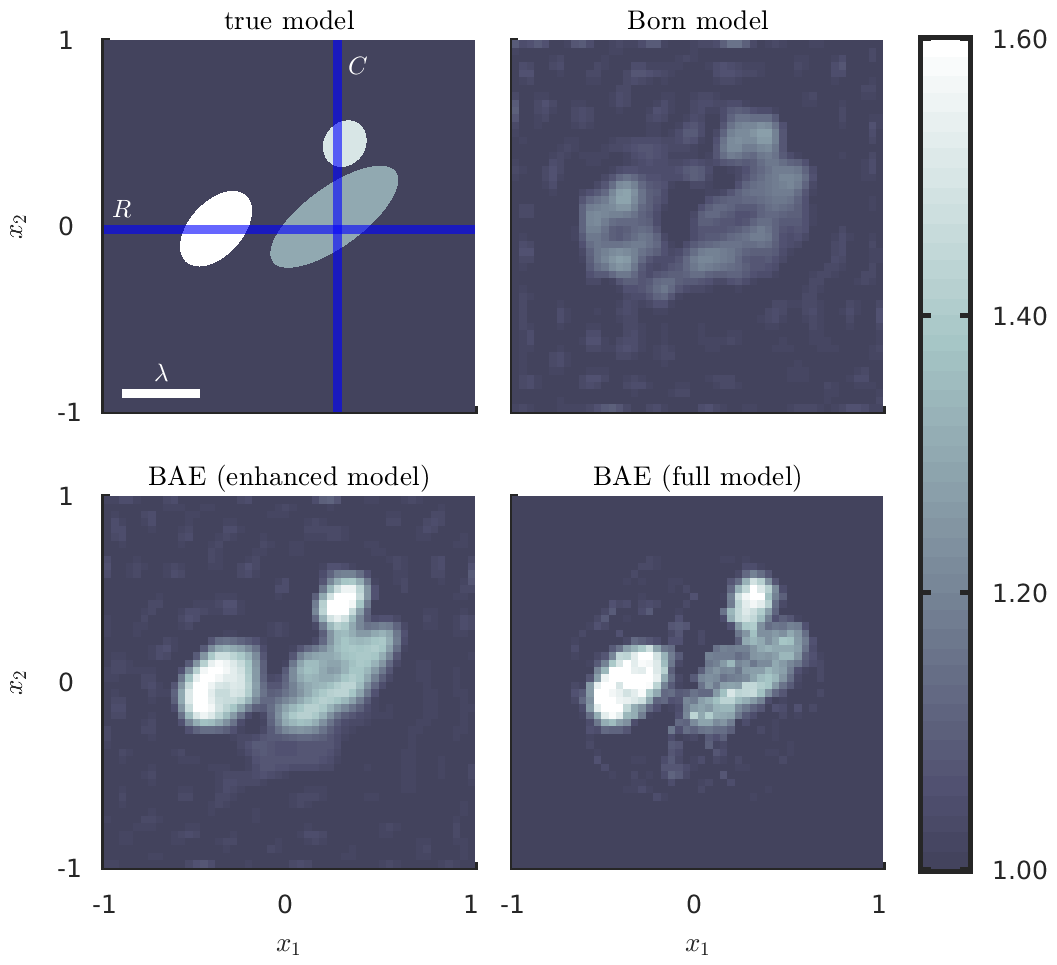}
 \end{center}
 \caption{The reconstruction of a three ellipsoidal scatterers with
   the $\bar n$ values ranging from 1.3 to 1.6. The panels are as in
   Fig.~\ref{fig:onescatterer_estimate}. Top left: true scatterer.
   Top right: Born approximation. Bottom left: enhanced error model.
   Bottom right: full BAE.  The actual $\bar n$ and the
   reconstructions along the cross sections $R$ and $C$ are shown in
   Fig.~\ref{fig:threescatterers_estimate_line}.}
   \label{fig:threescatterers_estimate}
\end{figure}

\begin{figure}[ht]
 \begin{center}
  \includegraphics[width=0.8\textwidth]{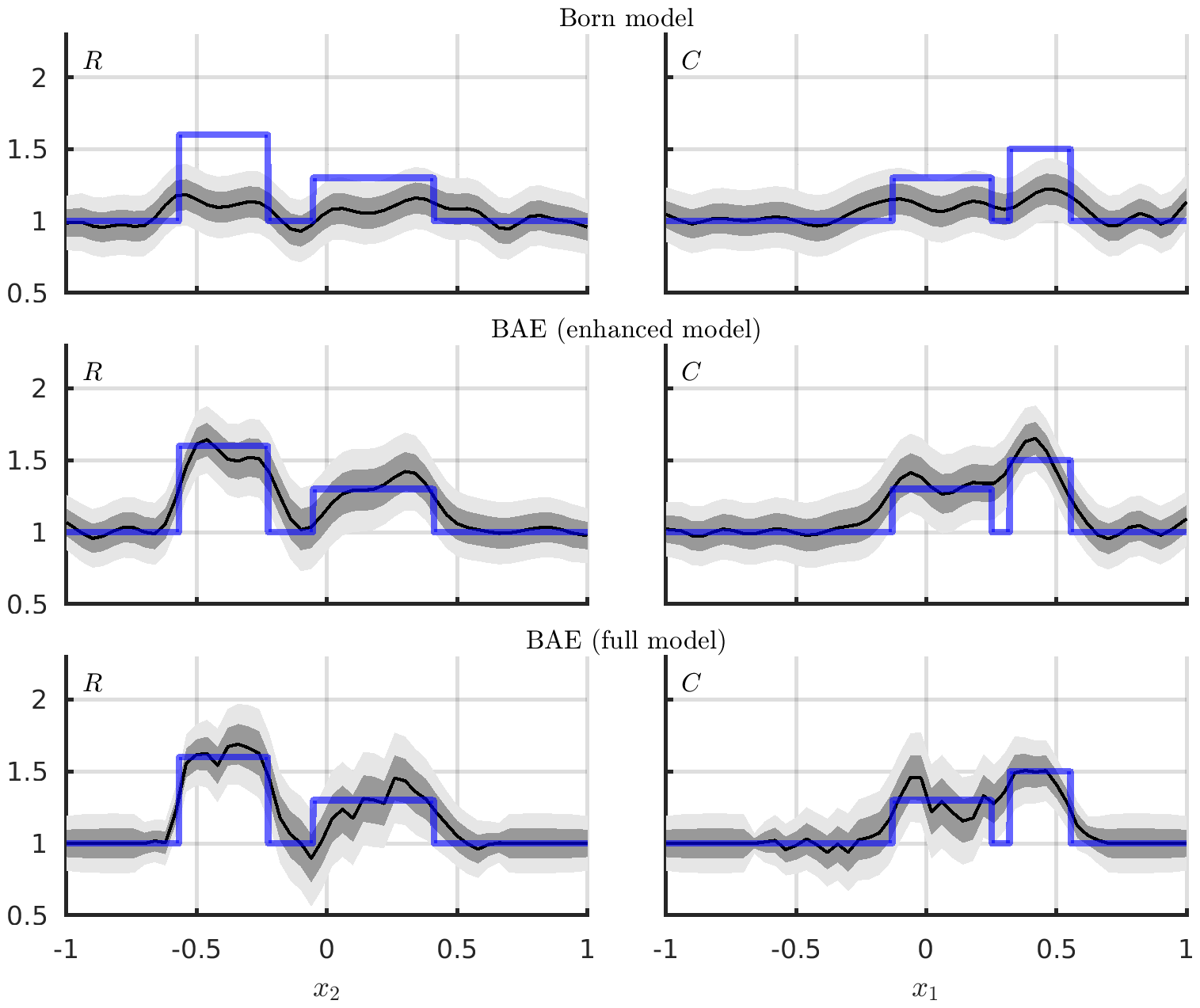}
 \end{center}
 \caption{The estimates for three elliptic scatterers along cross
   sections $R$ and $C$ of Fig. \ref{fig:threescatterers_estimate}.
   Left column: along $R$.  Right column: along $C$.}
 \label{fig:threescatterers_estimate_line}
\end{figure}

The second case contains three ellipsoidal scatterers for which the
$\bar n$ values range from 1.3 to 1.6.  The actual $\bar n$, the Born,
the full BAE and the enhanced error estimates are shown in
Fig.~\ref{fig:threescatterers_estimate}.  Qualitatively, the results
are similar to Case 1: The Born estimate has the general target
location correctly but the target values are badly off and shapes (and
the number of scatterers) are not recognisable.  Again, both BAE
estimates clearly exhibit the shapes, the values and the number
correctly.  Furthermore, in this case, the full error model is
slightly better in estimating the gaps between the scatterers.  Thus,
employing the BAE error model seems to make it possible to handle, to
a degree, also multiple scatterings between disjoint scatterers.  The
related cross sections are shown in
Fig.~\ref{fig:threescatterers_estimate_line}.  Again, the results are
qualitatively similar to Case 1.

\clearpage
%
\subsection{Case 3: A resonant structure}\label{case3}
\begin{figure}[ht]
 \begin{center}
  \includegraphics[width=0.8\textwidth]{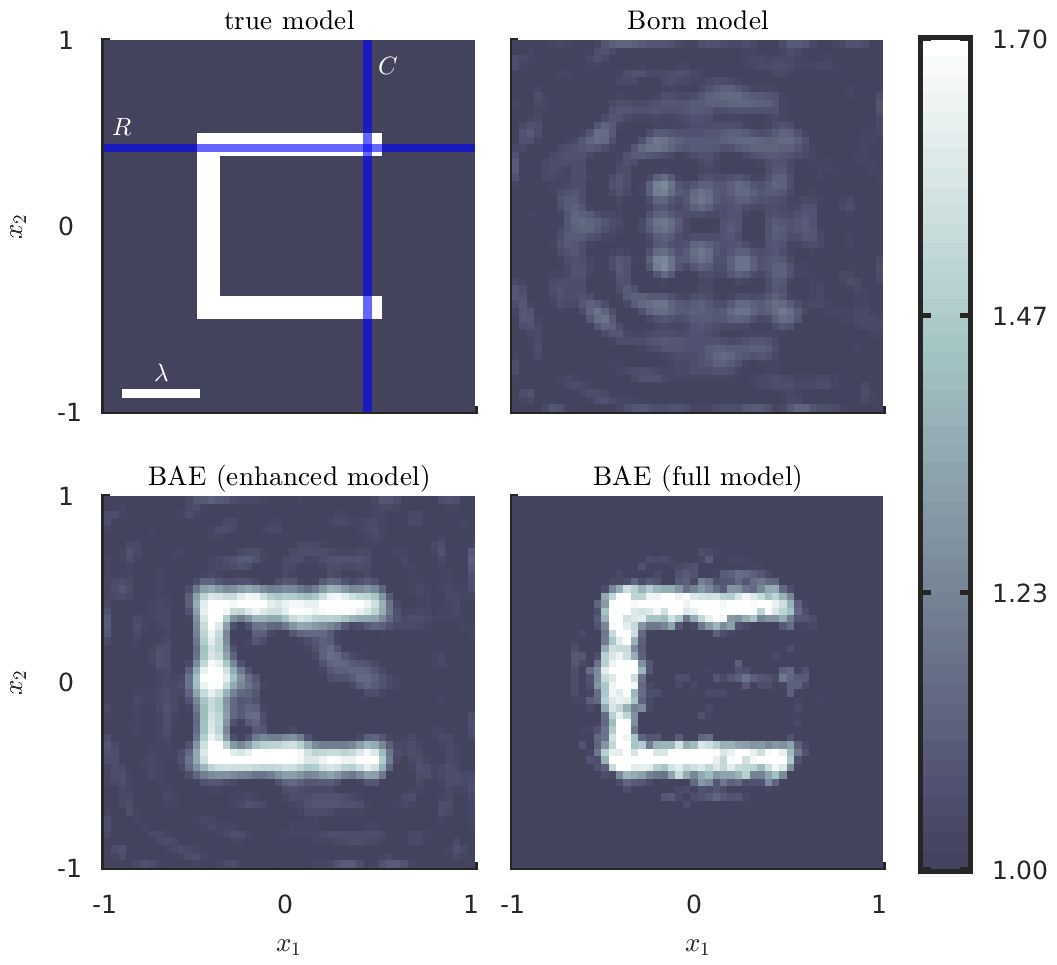}
 \end{center}
 \caption{The reconstruction of a high contrast ($\bar n=1.7$)
   resonant structure.  The layout of the figure is as in
   Fig.~\ref{fig:onescatterer_estimate}.  The actual $\bar n$ and the
   reconstructions along the cross sections $R$ and $C$ are shown in
   Fig.~\ref{fig:fork_estimate_line}. }
   \label{fig:fork_estimate}
\end{figure}

\begin{figure}[ht]
 \begin{center}
  \includegraphics[width=0.8\textwidth]{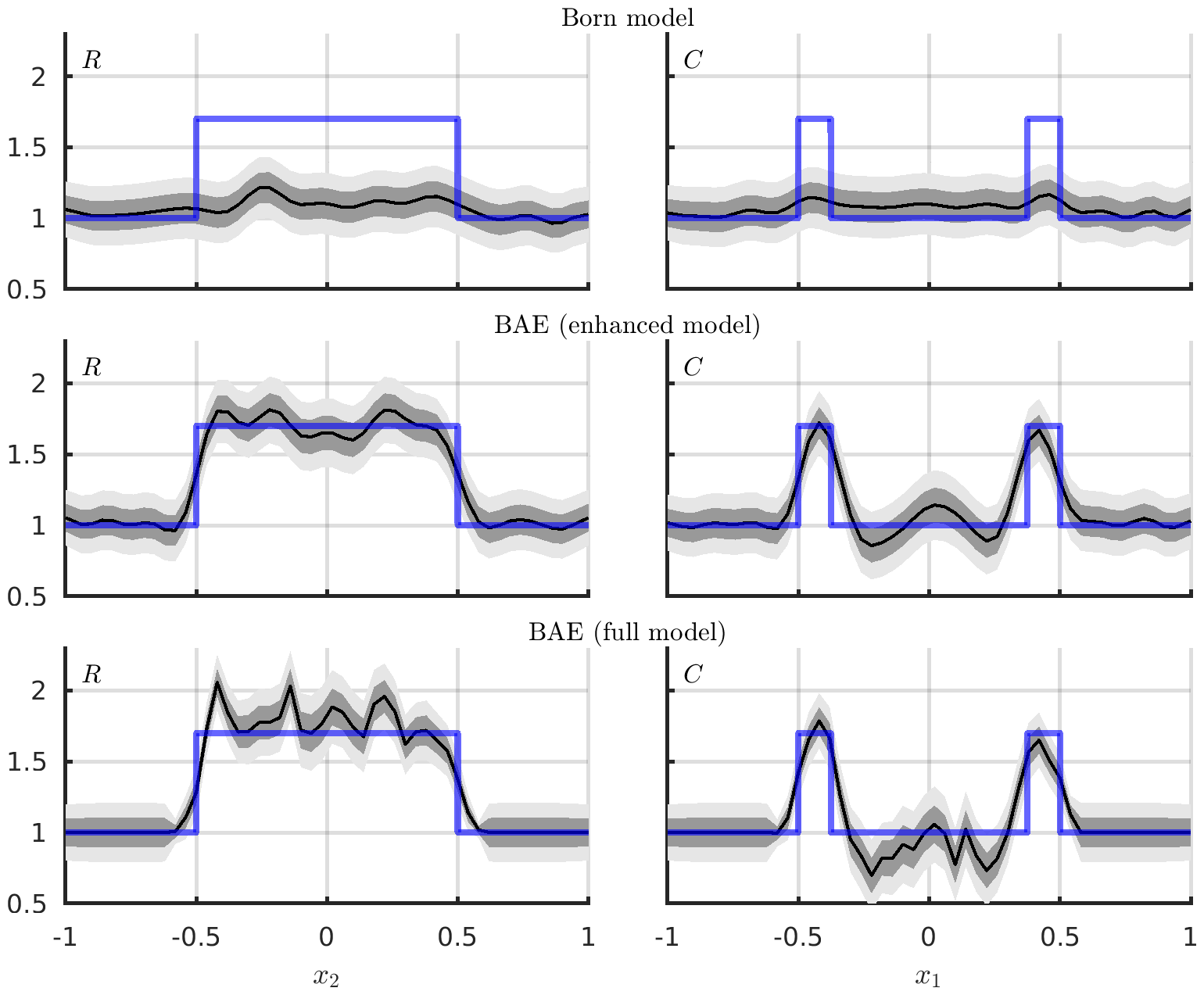}
 \end{center}
 \caption{The estimates of $n$ for the cavity along lines $R$ and $C$
   of Fig. \ref{fig:fork_estimate}. See
   Fig.~\ref{fig:onescatterer_estimate_line} for an explanation of
   each panel.}
   \label{fig:fork_estimate_line}
\end{figure}

The final target is a resonant structure shown in
Fig.~\ref{fig:fork_estimate}.  The walls of the structure have the
value $\bar n = 1.7$, this value, the shape and the wavelength make
the target a challenging case for the Born approximation, in
particular, due to the evident multiple scatterings.  While, in the
previous cases, the actual targets were consistent with the prior
model $\bar \pi(\bar n)$ (1-3 random ellipsoidal scatterers), this
example is not a feasible prior model.  Thus this example serves as a
test of the robustness of the approach.  The estimates and the
respective cross sections are shown in
Figs.~\ref{fig:fork_estimate}-\ref{fig:fork_estimate_line}.  In this
case, the Born estimate is completely off exhibiting mere ringing, as
was expected.  On the other hand, both BAE estimates get both the
shape and the values qualitatively correctly.  Furthermore, the BAE
results and their error estimates are useful, with the actual squared
refractive index generally lying within $\pm$2 standard deviations of
the reconstruction.
 
\section{Discussion}

In this paper, we considered the time harmonic inverse scattering
problem for a penetrable medium and the use of the Born approximation
with high contrast and multiple scattering targets.  It is well known
that the computationally appealing Born approximation is not accurate
for such scatterers.

We adopted the Bayesian approximation error approach to model the
error that is induced by employing the Born approximation as the
forward model.  This approach models the statistics of the difference
of the model predictions using an accurate and an approximate model.
This approximation error appears in the observation model as a
additive term that is correlated with the primary unknown.  By
carrying out a normal approximation to the joint distribution of the
primary unknown and the approximation errors, this term can be
marginalized analytically.  This procedure yields a computationally
efficient estimator that has the same or similar computational
complexity as the Born estimator under normal models and, for example,
a deterministic Tikhonov regularization.
 
The computational results suggest that the proposed approach can
handle high contrast and multiple scattering targets.  Most
importantly, the posterior error estimates are found to be useful: the
actual target typically lies within a few posterior standard deviation
intervals.

\section*{Acknowledgments}

This work has been supported by the Academy of Finland (Finnish Centre
of Excellence of Inverse Modelling and Imaging). The research of P.B.
Monk was partially supported by the US Air Force Office of Scientific
Research (AFOSR) under award number FA9550-17-1-0147.

\bibliographystyle{plain}

\end{document}